\documentclass{ws-procs9x6}

\begin{document}

\title{Baryon Resonances in the $1/{\rm N_c}$ Expansion}

\author{Richard F. Lebed}

\address{Department of Physics \& Astronomy, \\
Arizona State University, \\ 
Tempe, AZ 85287-1504, USA\\ 
E-mail: richard.lebed@asu.edu}

\maketitle

\abstracts{
The $1/N_c$ expansion of QCD provides a valuable semiquantitative tool
to study baryon scattering amplitudes and the short-lived baryon
resonances embedded within them.  A generalization of methods
originally applied in chiral soliton models in the 1980's provides the
key to deriving a rigorous $1/N_c$ expansion.  One obtains
model-independent relations among amplitudes that impose mass and
width degeneracies among resonances of various quantum numbers.
Phenomenological evidence confirms that patterns of resonant decay
predicted by $1/N_c$ agree with data.  One may extend the analysis to
subleading orders in $1/N_c$, where again agreement with data is
evident, in both meson-baryon scattering and pion photoproduction.}

\section{Introduction}

About 150 of the 1100 pages in the 2004 {\it Review of Particle
Properties}\cite{PDG} catalogue measured properties of baryons; and of
these, about 100 describe resonances unstable against strong decay,
with lifetimes so short as to appear only as features in partial wave
analyses.  Such states have resisted a model-independent description
for decades.  To date there exists no convincing explanation for why
QCD produces {\em any\/} baryon resonances, much less for their
peculiar observed spectroscopy, mass spacings, and decay widths.  Even
the unambiguous existence of numerous resonances remains open to
debate, as evidenced by the infamous 1- to 4-star classification
system.\cite{PDG}

Baryon resonances are exceptionally difficult to study precisely
because they {\em are\/} resonances rather than stable states.  For
example, treating baryon resonances as Hamiltonian eigenstates in
quark potential models is questionable, because such models are
strictly speaking valid only when vacuum $q \bar q$ pair production
and annihilation is suppressed (to ensure a Hermitian Hamiltonian).
It is just this mechanism, however, that provides the means by which
baryon resonances occur in scattering from ground-state baryons.

Even so, one of the most natural descriptions of excited baryons in
large $N_c$ remains an $N_c$ valence quark picture.  The inspiration
for this choice is that the ground-state baryon multiplets ($J^P
\! = \! {\frac 1 2}^+ \!$, ${\frac 3 2}^+$ for $N_c \! = \! 3$) neatly
fill a single multiplet completely symmetric under combined
spin-flavor symmetry [the SU(6) {\bf 56}, for 3 light flavors], so
that one may suppose the ground state of $N_c$ quarks is also
completely spin-flavor symmetric.  Indeed, the SU(6) spin-flavor
symmetry for ground-state baryons is shown to become exact in the
large $N_c$ limit.\cite{DM} Then, in analogy to the nuclear shell
model, excited states are formed by promoting a small number
[$O(N_c^0)$] of quarks into orbitally or radially excited orbitals.
For example, the generalization of the SU(6)$\times$O(3) multiplet
$({\bf 70}, 1^- )$ consists of $N_c - 1$ quarks in the ground state
and one in a relative $\ell \! = \! 1$ state.  One may then analyze
observables such as masses and axial-vector couplings by constructing
a Hamiltonian whose terms possess definite transformation properties
under the spin-flavor symmetry and are accompanied by known powers of
$N_c$.  By means of the Wigner-Eckart theorem, one then relates
observables for different states in each multiplet.  This approach has
been extensively studied\cite{CGKM,Goity,PY,CCGL,GSS,CC} (see
Ref.~\refcite{NStar} for a short review), but it falls short in two
important respects:

First, a Hamiltonian formalism is not entirely appropriate to unstable
particles, since it refers to matrix elements between asymptotic
external states.  Indeed, a resonance is properly represented by a
complex-valued pole in a scattering amplitude, its real and imaginary
parts indicating mass and width, respectively.  Moreover, a naive
Hamiltonian does not recognize the essential nature of resonances as
excitations of ground-state baryons.

Second, even a Hamiltonian constructed to respect the instability of
the resonances would not necessarily give states in the simple
quark-shell baryon multiplets as its eigenstates.  Just as in the
nuclear shell model, the possibility of {\it configuration mixing\/}
suggests that the true eigenstates might consist of mixtures of states
with 1, 2, or several excited quarks.

In contrast to quark potential models, chiral soliton models naturally
accommodate baryon resonances as excitations resulting from scattering
of mesons off ground-state baryons.  Such models are consistent with
the large $N_c$ limit because the solitons are heavy, semiclassical
objects compared to the mesons.  As has been known for many
years,\cite{ANW} a number of predictions following from the Skyrme and
other chiral soliton models are independent of the details of the
soliton structure, and may be interpreted as group-theoretical,
model-independent large $N_c$ results.  Indeed, the equivalence of
group-theoretical results for ground-state baryons in the Skyrme and
quark models in the large $N_c$ limit was demonstrated\cite{Manohar}
long ago.  Compared to quark models, chiral soliton models tend to
fall short in providing detailed spectroscopy and decay parameters for
baryon resonances, particularly at higher energies.  It is therefore
gratifying that large $N_c$ provides a point of reference where both
pictures share common ground.

In the remainder of this talk I discuss how the chiral soliton {\em
picture\/} (no specific model) may be used to study baryon resonances
as well as the full scattering amplitudes in which they appear, and
also its relation to the quark {\em picture\/} (again, no specific
model).  It summarizes a series of papers written in collaboration
with Tom Cohen (and more recently our
students),\cite{CL1,CLcompat,CDLN1,CLpent,CDLN2,CLSU3,CDLM} and
updates an earlier version\cite{UMN} of this talk.

\section{Amplitude Relations} \label{amp}

In the mid-1980's a series of papers\cite{HEHW,MK,MP,Mat3,MM}
uncovered a number of linear relations between meson-baryon scattering
amplitudes in chiral soliton models.  The fundamentally
group-theoretical nature of these results, as was pointed out,
suggested consistency with the large $N_c$ limit.

Standard $N_c$ counting\cite{Witten} shows that ground-state baryons
have masses of $O(N_c^1)$, but meson-baryon scattering amplitudes are
$O(N_c^0)$.  Therefore, the characteristic resonant energy of
excitation above the ground state and resonance widths are both
generically expected to be $O(N_c^0)$.  To say that two baryon
resonances are degenerate to leading order in $1/N_c$ thus actually
means equal masses at both the $O(N_c^1)$ and $O(N_c^0)$ levels.

A prototype of these linear relations was first derived in
Ref.~\refcite{MP}.  For a ground-state ($N$ or $\Delta$) baryon of
spin = isospin $R$ scattering with a meson (indicated by the
superscript) of relative orbital angular momentum $L$ (and primes for
analogous final-state quantum numbers) through a combined channel of
isospin $I$ and spin $J$, the full scattering amplitudes $S$ may be
expanded in terms of a smaller set of ``reduced'' scattering
amplitudes $s$:
\begin{eqnarray}
S_{LL^\prime R R^\prime IJ}^\pi & \! = & (-1)^{R^\prime \! \!
- R} \sqrt{[R][R^\prime]} \sum_K [K]
\left\{ \begin{array}{ccc} K &
I & J\\ R^\prime & L^\prime & 1 \end{array} \right\} \left\{
\begin{array}{ccc} K & I & J \\ R & L & 1 \end{array} \right\}
s_{K L^\prime L}^\pi \ , \ \label{MPeqn1} \\
S_{L R J}^\eta & = & \sum_K
\delta_{KL} \, \delta (L R J) \, s_{K}^\eta \ ,\label{MPeqn2}
\end{eqnarray}
where $[X] \equiv 2X \! + \! 1$, and $\delta(j_1 j_2 j_3)$ indicates
the angular momentum addition triangle rule.  Both are consequences of
a more general formula\cite{Mat9j} involving $9j$ symbols that holds
for mesons of arbitrary spin and isospin, which for brevity we do not
reproduce here.
%(See, {\it e.g.}, Ref.~\refcite{CLcompat}).
%
%\begin{eqnarray}
%S_{L_i L_f S_i S_f I J} & = & \sum_{K, \tilde{K}_i, \tilde{K}_f}
%[K] ([R_i][R_f][S_i][S_f][\tilde{K}_i][\tilde{K}_f])^{1/2}
%\nonumber \\ & &
%\times \left\{ \begin{array}{ccc}
%L_i & i_i & \tilde{K}_i \\
%S_i & R_i & s_i \\
%J & I & K \end{array} \right\} \left\{
%\begin{array}{ccc}
%L_f & i_f & \tilde{K}_f \\
%S_f & R_f & s_f \\
%J & I & K
%\end{array}
%\right\}
%\tau_{K \tilde{K}_i \tilde{K}_f L_i L_f},
%\label{Mmaster}
%\end{eqnarray}
%
%\footnote{Both are consequences of a
%more general formula\cite{Mat9j} involving $9j$ symbols that holds for
%mesons of arbitrary spin and isospin, which for brevity we decline to
%include.}
The basic feature inherited from chiral soliton models is the quantum
number $K$ ({\it grand spin}) with {\bf K}$\,\equiv\,${\bf
I}$\,$+$\,${\bf J}, conserved by the underlying hedgehog
configuration, which breaks $I$ and $J$ separately.  The physical
baryon state is a linear combination of $K$ eigenstates that is an
eigenstate of both $I$ and $J$ but no longer $K$.  $K$ is thus a good
(albeit hidden) quantum number that labels the reduced amplitudes $s$.
The dynamical content of relations such as
Eqs.~(\ref{MPeqn1})--(\ref{MPeqn2}) lies in the $s$ amplitudes, which
are independent for each value of $K$ allowed by $\delta (IJK)$.

In fact, $K$ conservation turns out to be equivalent to the large
$N_c$ limit.  The proof\cite{CL1} begins with the observation that the
leading-order (in $1/N_c$) $t$-channel exchanges have $I_t \! = \!
J_t,$\cite{KapSavMan} which in turn is proved using large $N_c$ {\em
consistency conditions}\cite{DJM}---essentially, unitarity
order-by-order in $1/N_c$ in meson-baryon scattering processes.
However, ($s$-channel) $K$ conservation was found---years earlier---to
be equivalent to the ($t$-channel) $I_t \! = \! J_t$ rule,\cite{MM}
due to the famous Biedenharn-Elliott sum rule,\cite{Edmonds} an SU(2)
identity.

The significance of Eqs.~(\ref{MPeqn1})--(\ref{MPeqn2}) lies in the
fact that there exist more full observable scattering amplitudes $S$
than reduced amplitudes $s$.  Therefore, one obtains a number of
linear relations among the measured amplitudes holding at leading
[$O(N_c^0)$] order.  In particular, a resonant pole appearing in one
of the physical amplitudes must appear in at least one reduced
amplitude; but this same reduced amplitude contributes to a number of
other physical amplitudes, implying a degeneracy between the masses
and widths of resonances in several channels.\cite{CL1} For example,
we apply Eqs.~(\ref{MPeqn1})--(\ref{MPeqn2}) to
negative-parity\footnote{Parity enters by restricting allowed values
of $L,L^\prime$.\cite{CLcompat}} $I \! = \! \frac 1 2$, $J \! = \frac
1 2$ and $\frac 3 2$ states (called $N_{1/2}$, $N_{3/2}$) in
Table~\ref{I}.  Noting that neither the orbital angular momenta $L ,
L^\prime$ nor the mesons $\pi , \eta$ that comprise the asymptotic
states can affect the compound state except by limiting available
total quantum numbers ($I$, $J$, $K$), one concludes that a resonance
in the $S_{11}^{\pi NN}$ channel ($K \!  = \! 1$) implies a degenerate
pole in $D_{13}^{\pi NN}$, because the latter contains a $K \! = \!
1$ amplitude.
\renewcommand\arraystretch{1.25}%
\begin{table}[ht]
\tbl{Application of Eqs.~(\ref{MPeqn1})--(\ref{MPeqn2}) to sample
negative-parity channels.
\label{I}}
{\footnotesize
\begin{tabular}{lcccccl}
\hline
State \mbox{ } && Quark Model Mass \mbox{ } &&
\multicolumn{3}{l}{Partial Wave, $K$-Amplitudes} \\
\hline
$N_{1/2}$ && $m_0$, $m_1$
   && $S^{\pi N N}_{11}$            &=& $s^\pi_{100}$ \\
&& && $D^{\pi \Delta \Delta}_{11}$  &=& $s^\pi_{122}$ \\
&& && $S^{\eta N N}_{1 1}$          &=& $s^\eta_0$ \\
\hline
%
%$\Delta_{1/2}$ && $m_1$, $m_2$
%   && $S^{\pi N N}_{31}$            &=& $s^\pi_{100}$ \\
%&& && $D_{31}^{\pi \Delta \Delta}$  &=& $\frac{1}{10}
%\left( s^\pi_{122} + 9 s^\pi_{222} \right)$ \\
%&& && $D^{\eta \Delta \Delta}_{31}$ &=& $s^\eta_2$ \\
%\hline
%
$N_{3/2}$ && $m_1$, $m_2$
   && $D^{\pi N N}_{13}$            &=& $\frac 1 2
\left( s^\pi_{122} + s^\pi_{222} \right)$ \\
&& && $D_{13}^{\pi N \Delta}$       &=& $\frac 1 2
\left( s^\pi_{122} - s^\pi_{222} \right)$ \\
&& && $S_{13}^{\pi \Delta \Delta}$  &=& $s^\pi_{100}$ \\
&& && $D_{13}^{\pi \Delta \Delta}$  &=& $\frac 1 2
\left( s^\pi_{122} + s^\pi_{222} \right)$ \\
&& && $D_{13}^{\eta N N}$           &=& $s^\eta_2$ \\
\hline
%
%$\Delta_{3/2}$  && $m_0$, $m_1$, $m_2$
%   && $D^{\pi N N}_{33}$            &=& $\frac{1}{20}
%\left( s^\pi_{122} + 5 s^\pi_{222} + 14 s^\pi_{322} \right)$ \\
%&& && $D^{\pi N \Delta}_{33}$ &=& $\frac{1}{5\sqrt{10}}
%\left( 2 s^\pi_{122} + 5 s^\pi_{222} - 7 s^\pi_{322} \right)$ \\
%&& && $S_{33}^{\pi \Delta \Delta}$  &=& $s^\pi_{100}$ \\
%&& && $D_{33}^{\pi \Delta \Delta}$  &=& $\frac{1}{25}
%\left( 8 s^\pi_{122} + 10 s^\pi_{222} + 7 s^\pi_{322} \right)$ \\
%&& && $S_{33}^{\eta \Delta \Delta}$ &=& $s^\eta_0$ \\
%&& && $D_{33}^{\eta \Delta \Delta}$ &=& $s^\eta_2$ \\
%\hline
%
%$N_{5/2}$ && $m_2$
%   && $D^{\pi N N}_{15}$            &=& $\frac{1}{9}
%\left( 2 s^\pi_{222} + 7 s^\pi_{322} \right)$ \\
%&& && $D_{15}^{\pi N \Delta}$       &=& $\frac{\sqrt{14}}{9}
%\left( s^\pi_{222} - s^\pi_{322} \right)$ \\
%&& && $D_{15}^{\pi \Delta \Delta}$  &=& $\frac{1}{9}
%\left( 7 s^\pi_{222} + 2 s^\pi_{322} \right)$ \\
%&& && $D_{15}^{\eta N N}$           &=& $s^\eta_2$ \\
%\hline
%
%$\Delta_{5/2}$ && $m_0$, $m_1$
%   && $D^{\pi N N}_{35}$            &=& $\frac{1}{90}
%\left( 27 s^\pi_{122} + 35 s^\pi_{222} + 28 s^\pi_{322} \right)$ \\
%&& && $D_{35}^{\pi N\Delta}$        &=&
%$\frac{1}{90} \sqrt{\frac{7}{5}}
%\left( 27 s^\pi_{122} + 5 s^\pi_{222} - 32 s^\pi_{322} \right)$ \\
%&& && $D_{35}^{\pi \Delta \Delta}$  &=& $\frac{1}{450}
%\left( 189 s^\pi_{122} + 5 s^\pi_{222} + 256 s^\pi_{322} \right)$ \\
%&& && $D_{35}^{\eta \Delta \Delta}$ &=& $s^\eta_2$ \\
%\hline
%
%$\Delta_{7/2}$ && $m_2$
%&& $D_{37}^{\pi \Delta \Delta}$ &=& $\frac 1 5
%\left( 2 s^\pi_{222} + 3 s^\pi_{322} \right)$ \\
%\hline
\end{tabular} }
\end{table}
\renewcommand\arraystretch{1.0}%
One thus obtains towers of degenerate negative-parity resonance
multiplets labeled by $K$:
%
%{\footnotesize
\begin{eqnarray}
N_{1/2} , \; \Delta_{3/2} , \; \cdots \; &~& (K \! = \! 0 \! : \,
s_{0}^\eta) \; , \nonumber \\
N_{1/2} , \; \Delta_{1/2} , \; N_{3/2} , \; \Delta_{3/2} , \;
\Delta_{5/2} , \; \cdots \; &~& (K \! = \! 1 \! :
\, s_{1 0 0}^\pi , \, s_{1 2 2}^\pi) \; , \nonumber \\
\Delta_{1/2} , \; N_{3/2} , \; \Delta_{3/2} , \; N_{5/2} , \;
\Delta_{5/2} , \; \Delta_{7/2} , \; \cdots \;
&~& (K \! = \! 2 \! : \, s_{2 2 2}^\pi, \, s_{2}^\eta ) \; .
%\nonumber \\
%\Delta_{3/2} , \; N_{5/2} , \; \Delta_{5/2} , \; \Delta_{7/2} , \;
%\cdots \; &~& (s^\pi_{3 2 2} )
\label{towers}
\end{eqnarray}
%}
%
%where the states are listed on the left and the contributing
%amplitudes on the right.  The ellipses indicate that in the large
%$N_c$ world the multiplets are infinite dimensional, and we have
%simply listed the low-spin and -isospin members of the multiplet.

It is now fruitful to consider the quark-shell picture large $N_c$
analogue of the first excited negative-parity multiplet [the $({\bf
70}, 1^- )$].  Just as for $N_c \! = \! 3$, there are two $N_{1/2}$
and two $N_{3/2}$ states.  If one computes the masses to $O(N_c^0)$
for the entire multiplet in which these states appear, one finds only
three distinct eigenvalues,\cite{CCGL,CL1,PS} which are labeled $m_0$,
$m_1$, and $m_2$ and listed in Table~\ref{I}.  Upon examining an
analogous table containing all the states in this multiplet,\cite{CL1}
one quickly concludes that exactly the required resonant poles are
obtained if each $K$ amplitude, $K \! = \! 0,1,2$, contains precisely
one pole, which is located at the value $m_K$.  The lowest quark-shell
multiplet of negative-parity excited baryons is found to be {\em
compatible\/} with, {\it i.e.}, consist of a complete set of,
multiplets classified by $K$.  But the quark-shell masses are {\em
real\/} Hamiltonian eigenvalues, and therefore present a result less
general than that obtained from the $K$ amplitude analysis.

One can prove\cite{CLcompat} this compatibility for all nonstrange
baryon multiplets in the SU(6)$\times$O(3) shell
picture.\footnote{Studies to extend these results to flavor SU(3) are
underway\cite{CLSU3}; while the group theory is more complicated, it
remains tractable.}  It is important to note that compatibility does
not imply SU(6) is an exact symmetry at large $N_c$ for resonances as
it is for ground states.\cite{DM} Instead, it says that
SU(6)$\times$O(3) multiplets are {\em reducible\/} multiplets at large
$N_c$.  In the example given above, $m_{0,1,2}$ each lie only
$O(N_c^0)$ above the ground state, but are separated by $O(N_c^0)$
intervals.

We emphasize that large $N_c$ by itself does not mandate the existence
of any resonances at all; rather, it merely tells us that if even one
exists, it must be a member of a well-defined multiplet.  Although the
soliton and quark pictures both have well-defined large $N_c$ limits,
compatibility is a remarkable feature that combines them in a
particularly elegant fashion.

\section{Phenomenology} \label{phenom}

Confronting these formal large $N_c$ results with experiment poses two
significant challenges, both of which originate from neglecting
$O(1/N_c)$ corrections.  First, the lowest multiplet of nonstrange
negative-parity states covers quite a small mass range (only
1535--1700~MeV), while $O(1/N_c)$ mass splittings can generically be
as large as $O$(100~MeV).  Any claims that two such states are
degenerate while two others are not must be carefully scrutinized.
Second, the number of states in each multiplet increases with $N_c$,
meaning that a number of large $N_c$ states are spurious in $N_c \! =
\! 3$ phenomenology.  For example, for $N_c \! \ge \! 7$ the analogue
of the {\bf 70} contains three $\Delta_{3/2}$ states, but only one
[$\Delta(1700)$] when $N_c \! = \! 3$.  As $N_c$ is tuned down from
large values toward 3, the spurious states must decouple through the
appearance of factors such as $(1-3/N_c)$, which in turn requires one
to understand simultaneously leading and subleading terms in the
$1/N_c$ expansion.

Nevertheless, it is possible to obtain testable predictions for the
decay channels, even using just the leading-order results.  For
example, note from Table~\ref{I} that the $K \! = \! 0(1)$ $N_{1/2}$
resonance couples only to $\eta$($\pi$).  Indeed, the $N(1535)$
resonance decays to $\eta N$ 30--55$\%$ of the time despite lying
barely above that threshold, while the $N(1650)$ decays to $\eta N$
only 3--10$\%$ of the time despite having much more comparable phase
space to $\pi N$ and $\eta N$.  This pattern clearly suggests that the
$\pi$-phobic $N(1535)$ should be identified with $K \! = \!  0$ and
the $\eta$-phobic $N(1650)$ with $K \! = \! 1$, the first fully field
theory-based explanation for these phenomenological facts.

\section{Configuration Mixing}

As mentioned above, one does not expect quark-shell baryon states with
a fixed number of excited quarks to be eigenstates of the full QCD
Hamiltonian.  Rather, configuration mixing likely clouds the
situation.  Consider, for example, the expectation that baryon
resonances have generically broad [$O(N_c^0)$] widths.  One may ask
whether some states escape this restriction and turn out to be narrow
in the large $N_c$ limit.  Indeed, some of the first work\cite{PY} on
excited baryons combined large $N_c$ consistency conditions and a
quark description of excited baryon states to predict that baryons in
the {\bf 70}-analogue have widths of $O(1/N_c)$, while states in an
excited negative-parity spin-flavor symmetric multiplet ({\bf
56$^\prime$}) have $O(N_c^0)$ widths.

In fact there arise, even in the quark-shell picture, operators
inducing configuration mixing between these multiplets.\cite{CDLN1}
The spin-orbit and spin-flavor tensor operators (respectively $\ell s$
and $\ell^{(2)} g \, G_c$ in the notation of
Refs.~\refcite{CCGL},$\,$\refcite{GSS},$\,$\refcite{PS}), which appear
at $O(N_c^0)$ and are responsible for splitting the eigenvalues $m_0$,
$m_1$, and $m_2$, give nonvanishing transition matrix elements between
the {\bf 70} and {\bf 56$^\prime$}.  Since states in the latter
multiplet are broad, configuration mixing forces at least some states
in the former multiplet to be broad as well.  One concludes that the
possible existence of any excited baryon state narrow in the large
$N_c$ limit requires a fortuitous absence of significant configuration
mixing.

\section{Pentaquarks}

The possible existence of a narrow isosinglet, strangeness +1 (and
therefore exotic) baryon state $\Theta^+ (1540)$, claimed to be
observed by numerous experimental groups (but not seen by several
others), remains an issue of great dispute.  Although the jury remains
out on this important question, one may nevertheless use the large
$N_c$ method described above to determine the quantum numbers of its
degenerate partners.\cite{CLpent} For example, if one imposes the
theoretical prejudice $J_{\Theta} \! = \! \frac 1 2$, then there must
also be pentaquark states with $I \! = \! 1$, $J \! = \! \frac 1 2,
\frac 3 2$ and $I \! = \! 2$, $J \! = \! \frac 3 2, \frac 5 2$, with
masses and widths equal that of the $\Theta^+$, up to $O(1/N_c)$
corrections.

The large $N_c$ analogue of the ``pentaquark'' actually carries the
quantum numbers of $N_c \! + \! 2$ quarks, consisting of $(N_c \!  +
\! 1)/2$ spin-singlet, isosinglet $ud$ pairs and an $\bar s$ quark.
The quark operator picture, for example, shows the partner states we
predict to belong to SU(3) multiplets {\bf 27} ($I \! = \! 1$) and
{\bf 35} ($I \! = \! 2$).\cite{JMpent} However, the existence of
partners does not depend upon any particular picture for the resonance
or any assumptions regarding configuration mixing.  Since the generic
width for such baryon resonances remains $O(N_c^0)$, the surprisingly
small reported width ($<$10 MeV) does not appear to be explicable by
large $N_c$ considerations alone, but may be a convergence of small
phase space and a small nonexotic-exotic-pion coupling.

\section{$1/N_c$ Corrections}

All the results exhibited thus far hold at the leading nontrivial
order ($N_c^0$) in the $1/N_c$ expansion.  We saw in Sec.~\ref{phenom}
that $1/N_c$ corrections are essential not only to explain the sizes
of effects apparent in the data, but in the very enumeration of
physical states.  Clearly, if this analysis is to carry real
phenomenological weight, one must demonstrate a clear path to
characterize $1/N_c$ corrections to the scattering amplitudes.
Fortunately, such a generalization is possible: As discussed in
Sec.~\ref{amp}, the constraints on scattering amplitudes obtained from
the large $N_c$ limit are equivalent to the $t$-channel requirement
$I_t \! = \! J_t$.  In fact, Refs.~\refcite{KapSavMan} showed not only
that the large $N_c$ limit imposes this constraint, but also that
exchanges with $|I_t \! - \! J_t| \! = \! n$ are suppressed by a
relative factor $1/N_c^n$.

This result permits one to obtain relations for the scattering
amplitudes incorporating all effects up to and including $O(1/N_c)$:
\begin{eqnarray}
\! \! \! \! S_{LL^\prime R R^\prime I_s J_s} \! \! \! \! & = &
\sum_{\mathcal J} \left[
\begin{array}{ccc} 1 & R^\prime & I_s \\ R & 1 & I_t \! = \!
{\mathcal J}
\end{array} \right] \left[
\begin{array}{ccc} L^\prime & R^\prime & J_s \\ R & L & J_t \! = \!
{\mathcal J}
\end{array}
\right] s_{{\mathcal J} L L^\prime}^t \nonumber \\ &
-\frac{1}{N_c} & \sum_{\mathcal J} \left[ \begin{array}{ccc} 1 &
R^\prime & I_s\\ R & 1 & I_t \! = \! {\mathcal J}
\end{array} \right] \left[
\begin{array}{ccc} L^\prime & R^\prime & J_s \\ R & L &
J_t \! = \! {{\mathcal J} \! + \! 1}
\end{array}
\right] s_{{\mathcal J} L L^\prime}^{t(+)} \nonumber \\ &
- \frac{1}{N_c} & \sum_{\mathcal J} \left[
\begin{array}{ccc} 1 & R^\prime & I_s\\ R & 1 & I_t \! = \!
{\mathcal J}
\end{array} \right] \left[
\begin{array}{ccc} L^\prime & R^\prime & J_s \\ R & L &
J_t \! = \! {{\mathcal J} \! - \! 1}
\end{array}
\right] s_{{\mathcal J} L L^\prime}^{t(-)} + 
O(\mbox{\small{$\frac{1}{N_c^2}$}}) ,
\label{MPplus}
\end{eqnarray}
One obtains this expression by first rewriting $s$-channel expressions
such as Eqs.~(\ref{MPeqn1})--(\ref{MPeqn2}) in terms of $t$-channel
amplitudes.  The $6j$ symbols in this case contain $I_t$ and $J_t$ as
arguments (which for the leading term are equal).  One then
introduces\cite{CDLN2} new $O(1/N_c)$-suppressed amplitudes
$s^{t(\pm)}$, for which $J_t - I_t = \pm 1$.  The square-bracketed $6j$
symbols in Eq.~(\ref{MPplus}) differ from the usual ones only through
normalization factors, and in particular obey the same triangle rules.

\begin{figure}[ht]
%\epsfxsize=10cm   %width of figure - will enlarge/reduce the figures
%\epsfbox{fig3.eps}
%\figurebox{2cm}{3cm}{} %to have a box alone 
\centerline{\epsfxsize=4.1in\epsfbox{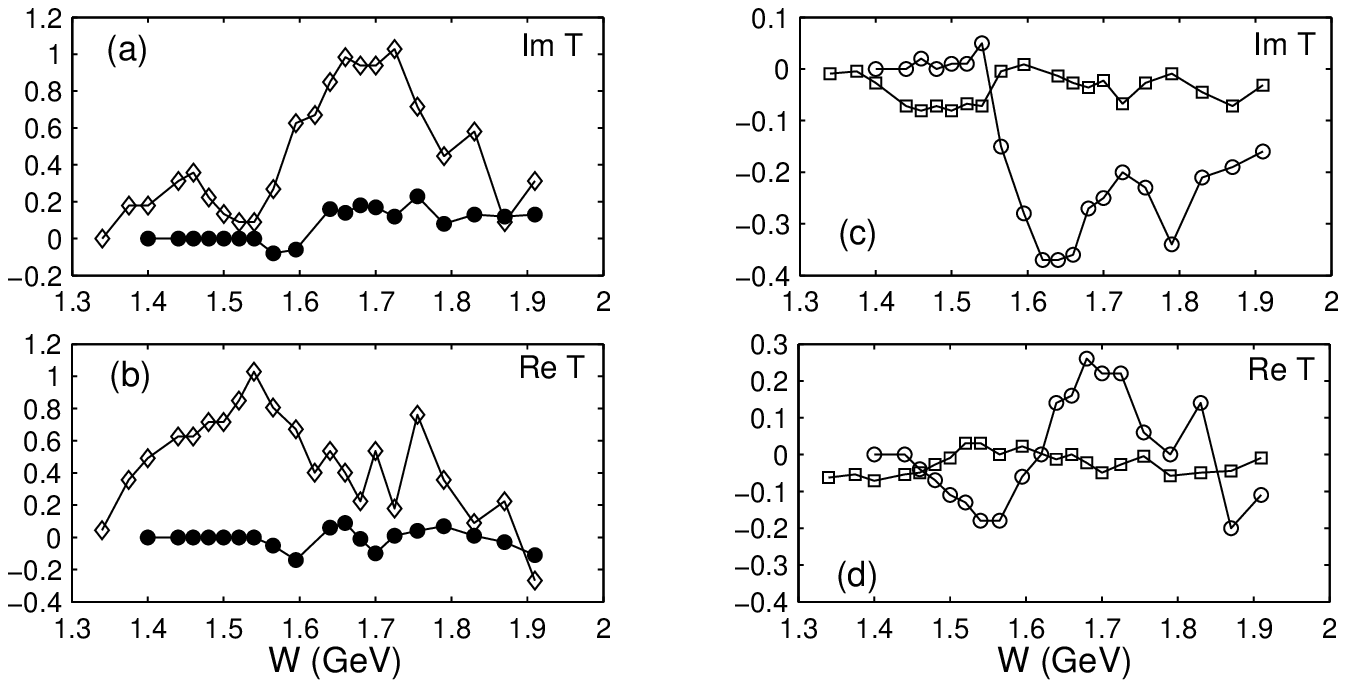}}
\centerline{\epsfxsize=4.1in\epsfbox{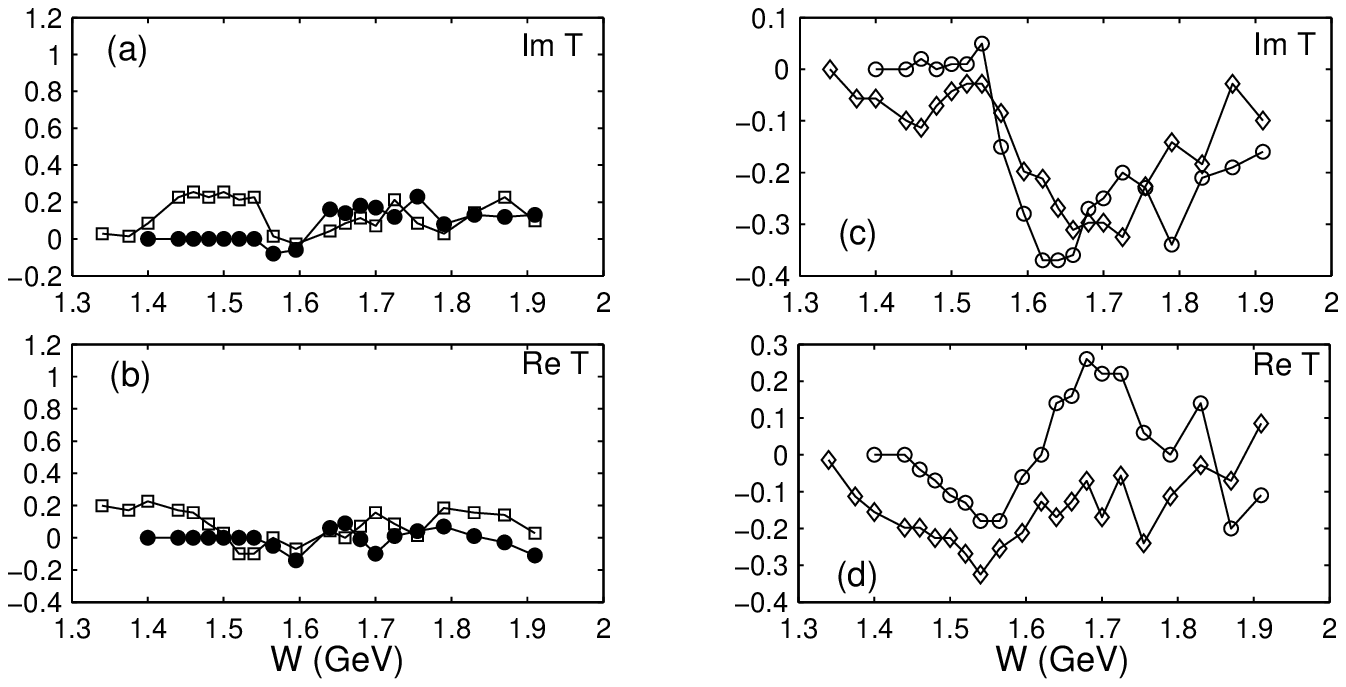}}   
\caption{Real and imaginary parts of $\pi N \to \pi \Delta$ scattering
amplitudes.  The first four insets give two particular partial waves
equal to leading order [hence indicating the size of $O(1/N_c)$
corrections].  The second four insets give two particular linear
combinations of the same data good to $O(1/N_c^2).$\label{inter}}
\end{figure}

Relations between observable amplitudes that incorporate the larger
set $s^t$, $s^{t(+)}$, and $s^{t(-)}$ are expected to be a factor of
$N_c \! = \! 3$ better than those merely including the leading
$O(N_c^0)$ results.  Indeed, this is dramatically evident in $\pi N \!
\to \! \pi \Delta$, where sufficient numbers of amplitudes are
measured (Fig.~\ref{inter}).  For example, (c) and (d) in the first
four insets give the imaginary and real parts, respectively, of
partial wave data for $SD_{31}$ ({\Large\rm$\circ$}) and $(1/\sqrt{5})
DS_{13}$ ($\square$), which are equal up to $O(1/N_c)$ corrections; in
(c) and (d) of the second four insets, the {\Large\rm $\circ$} points
again are $SD_{31}$ data, while $\lozenge$ represent $-\sqrt{2}
DS_{33}$, and by Eq.~(\ref{MPplus}) these are equal up to $O(1/N_c^2)$
corrections.

\section{Pion Photoproduction}

Meson-baryon scattering is not the only process that can be considered
in the soliton-inspired picture.  As long as one knows the isospin and
spin quantum numbers of the field coupling to the baryon along with
the corresponding $N_c$ power suppression of each coupling, one may
carry out precisely the same sort of analysis as described above.

The processes we have in mind are those involving real or virtual
photons (photoproduction,\cite{CDLM} electroproduction, real or
virtual Compton scattering).  One minor complication is that the
electromagnetic interaction breaks isospin, in that the photon is a
mixture of isoscalar ($I \! = \! 0$) and isovector ($I \!  = \! 1$)
sources.  The former is suppressed by a factor $1/N_c$ compared to the
latter since baryon couplings carrying both a spin index (coupling to
the photon polarization vector) and an isospin index are larger than
those carrying just a spin index by a factor $N_c$.\cite{JJM}

Moreover, electromagnetic processes are typically parametrized in
terms of multipole amplitudes, which combine the intrinsic photon spin
with its relative orbital angular momentum; in fact, this is very
convenient, because then the photon can be treated effectively as a
spinless field whose effective orbital angular momentum is the order
of the multipole.  Note that this makes processes with virtual photons
just as simple as those with real photons, even though the former can
carry not only spin-1 but spin-0 amplitudes as well.  With these
caveats in mind, carrying out an analysis of pion photoproduction
amplitudes, including $1/N_c$ corrections (leading plus subleading $I
\! = \! 1$ amplitudes and leading $I \! = \! 0$ amplitudes), is
straightforward.\cite{CDLM}

For example, a relationship receiving only $O(1/N_c^2)$ corrections
reads
\begin{equation}
M^{\textrm{m},\,p(\pi^+)n}_{L,L,-}
=
M^{\textrm{m},\,n(\pi^-)p}_{L,L,-}
-\left(\frac{L+1}{L}\right)
\left[
M^{\textrm{m},\,p(\pi^+)n}_{L,L,+}-M^{\textrm{m},\,n(\pi^-)p}_{L,L,+}
\right] , \label{NLO}
\end{equation}
where the superscript m means magnetic multipoles, $N (\pi^a)
N^\prime$ means the process $N \gamma \! \to \! N^\prime \pi^a$, and
the subscripts $L, L, \pm$ mean that an electromagnetic multipole of
order $L$ creates a pion in the $L^{\rm th}$ partial wave, with total
$J \! = \!  L \! \pm \! \frac 1 2$.  Including just the first term on
the right-hand side (r.h.s.)  gives a relation valid up to $O(1/N_c)$
corrections, and the quality of both this relation and its extension
to next-to-leading order may be assessed.

\begin{figure}[ht]
%\epsfxsize=10cm   %width of figure - will enlarge/reduce the figures
%\figurebox{1.25in}{4.5in}{} %to have a box alone
\epsfxsize 2.2 in \epsfbox{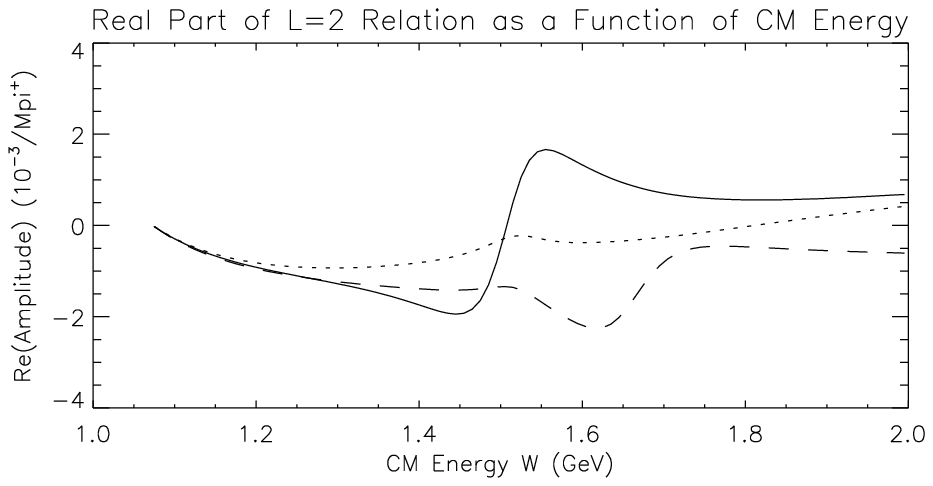} \hspace{0em}
\epsfxsize 2.2 in \epsfbox{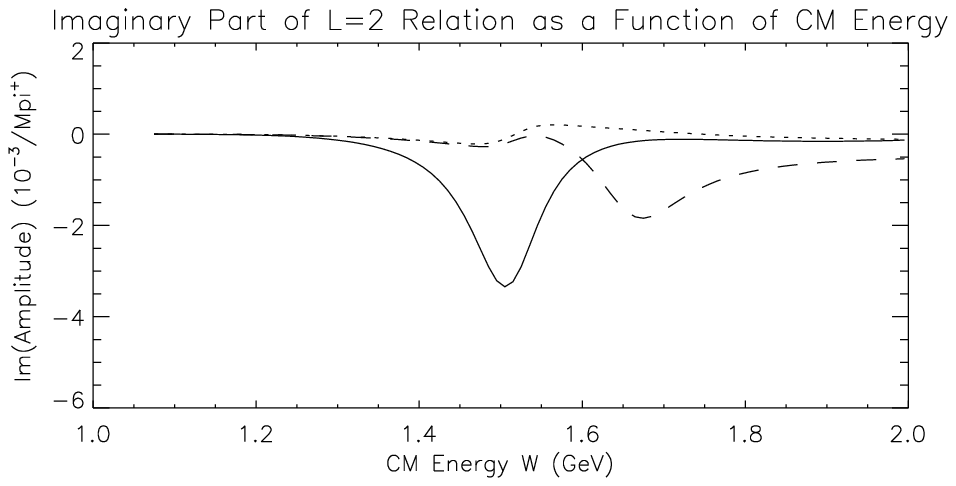} \\
%\centerline{\epsfxsize=4.1in\epsfbox{L_2_Real_MAID.eps}}
%\centerline{\epsfxsize=4.1in\epsfbox{L_2_Imag_MAID.eps}}   
\vskip -1.0ex
\caption{Comparison of Eq.~(\ref{NLO}) with data for $L \! = \! 2$.  The
solid line is the l.h.s., the dotted line is the first term on the
r.h.s., and the dashed line is all r.h.s.\ terms.\label{MAID}}
\end{figure}

A sample result appears in Fig.~\ref{MAID}, where the left-hand side
(l.h.s.)  is a solid line, the $O(1/N_c)$ result is dotted, and the
$O(1/N_c^2)$ is dashed.  While the agreement at first glance may not
seem impressive, some very heartening features may be discerned.
First, the agreement in the region below the appearance of resonances
is quite good, and indeed improves at $O(1/N_c^2)$.  Second, unlike
the solid line [containing D$_{13}$(1520)], the dotted line gives no
hint of a resonance but the dashed line does [D$_{15}$(1675)]; and the
fact that their positions do not precisely match should not alarm us,
as one expects them to differ by an amount of $O(\Lambda_{\rm
QCD}/N_c) \! \sim \! 100$~MeV.  One may in fact use the helicity
amplitudes compiled\cite{PDG} for these two resonances and relate them
directly to the amplitudes appearing in Eq.~(\ref{NLO}).  In order to
obtain dimensionless and scale-independent results, one divides the
linear combination of helicity amplitudes corresponding to
Eq.~(\ref{NLO}) by the same expression with all signs made positive.
The $O(1/N_c)$ and $O(1/N_c^2)$ combinations give\cite{CDLM} $-0.38 \!
\pm \!  0.06$ and $-0.13 \pm 0.06$, respectively, showing that the
$1/N_c$ expansion works beautifully---even better than one might
expect.

\section{Conclusions: The Way Forward}

There now exist reliable and convincing calculational techniques using
the $1/N_c$ expansion of QCD that handle not only long-lived
ground-state baryons, but also unstable baryon resonances and the
scattering amplitudes in which they appear.  The approach, originally
noted in chiral soliton models but eventually shown to be a true
consequence of large $N_c$ QCD, is found to have phenomenological
consequences [such as the large $\eta$ branching fraction of the
$N(1535)$] that compare favorably with real data.

The first steps of obtaining $1/N_c$ corrections to the leading-order
results, absolutely essential to make comparisons with the full data
set, are complete.  The measured scattering amplitudes appear to obey
the constraints placed by these corrections, and more work along these
lines is forthcoming.  For example, the means by which the spurious
extra resonances of $N_c \! > \! 3$ decouple as one takes the limit
$N_c \! \to \! 3$ is crucial and not yet understood.

The explicit results presented here, as mentioned in Sec.~\ref{amp},
have used only relations among states of fixed strangeness.  Moving
beyond this limitation means using flavor SU(3) group theory, which is
rather more complicated than isospin SU(2) group theory.
Nevertheless, this is merely a technical complication, and existing
work shows that it can be overcome.\cite{GSS,CLSU3}

At the time of this writing, all of the essential tools appear to be
in place to commence a full-scale analysis of baryon scattering and
resonance parameters.  One may envision a sort of resonance
calculation factory, which I have previously dubbed {\em Baryons
$IN_C$}.\cite{UMN} Given sufficient time and researchers, the whole
baryon resonance spectrum can be disentangled using a solid,
field-theoretical approach based upon a well-defined limit of QCD.
\vspace{-1.5ex}

\section*{Acknowledgments}

I thank my co-organizers for continuing the tradition of large $N_c$
meetings and for assembling a excellent program.  I also appreciate
the fine assistance of the ECT$^*$ staff.  Finally, I would like to
thank Korova Coffee Bar, San Diego, whose superb French Roast and free
wireless internet access aided the timely writing of this paper.  The
work described here was supported in part by the National Science
Foundation under Grant No.\ PHY-0140362.

\end{document}